# DESIGN AND IMPLEMENTATION OF CAR PARKING SYSTEM ON FPGA

Ramneet Kaur[1] and Balwinder Singh[2]

[1,2]Academic and Consultancy Services-Division, Centre for Development of Advanced Computing(C-DAC), Mohali, India
`romy.grewal17@gmail.com, balwinder.cdacmohali@gmail.com`

## ABSTRACT

*As, the number of vehicles are increased day by day in rapid manner. It causes the problem of traffic congestion, pollution (noise and air). To overcome this problem A FPGA based parking system has been proposed. In this paper, parking system is implemented using Finite State Machine modelling. The system has two main modules i.e. identification module and slot checking module. Identification module identifies the visitor. Slot checking module checks the slot status. These modules are modelled in HDL and implemented on FPGA. A prototype of parking system is designed with various interfaces like sensor interfacing, stepper motor and LCD.*

## KEYWORDS

*Finite State Machine; Parking System; Virtex- 5;*

## 1. INTRODUCTION

Vehicle traffic congestion is a worldwide problem.  In recent years, efforts have been made to introduce a method to reduce parking problems such as congestion, accidents and hazards.

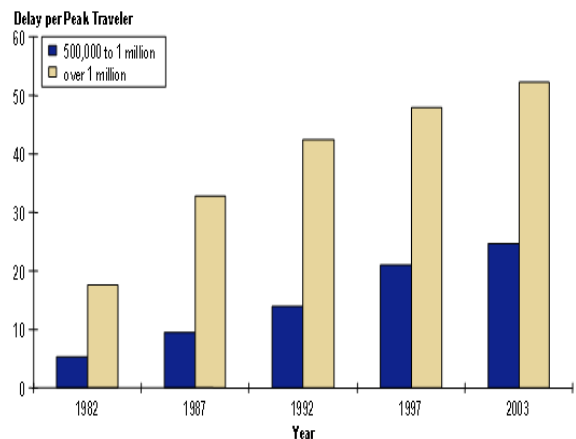

Figure 1: Congestion Trends in Urban Area [11]

                                                                                                                 

As shown in figure 1 congestion has clearly grown year by year. It creates a number of problems. Congestion used to mean it took longer to get to/ from work in the "rush hour"[11]. Parking systems can also take advantage of innovative technologies in order to improve the ease and convenience of paying for parking. Now a day, Smart cards minimize transaction time by allowing a user to simply wave their card in front of a reader. Mobile devices can also be used in payment transactions. Public utilities need a parking system that can function efficiently and be integrated with the other urban city utilities. For allotment of parking slots there is no proper way thus parking management system fails in coordination and centralizing the information for an effective system. To avoid these problems, a design of an intelligent parking system is proposed, which will be implemented on FPGA to check its functionality.

Recently, a reconfigurable FPGA is efficient method to implement a design, because FPGA provides a compromise between general-purpose processors and ASIC. The FPGA based design is also more flexible, programmable and can be re-programmed. FPGA based design can easily be modified by modifying design's software part.

## 2. RELATED WORK

Gongjun Yan et.al, (2011) describes a novel, secure, and intelligent parking system (Smart Parking) based on secured wireless network and sensor communication [2]. High parking space utilization and fast free spot finding time are the result of proposed research. Soh Chun Khang et.al, (2010) presents a parking system in which driver comes to know about the space availability in the parking lot with the help of SMS service. Driver can resend SMS in order to request new space if the previous one is filled. Driver can find nearest space for parking using wireless mobile based car parking system. Results, shows that the system efficiently allocates the slots and utilizes the full parking space [4]. Ankit Gupta et.al, (2010) describes an efficient car parking algorithm for ackerman steering configuration. This algorithm uses geometric calculations for path planning. Result shows a fast, efficient and safer parking system [5]. Hua-chun tan et.al, (2009) proposed an efficient car searching technique for larger parking lot. In this paper, cameras are installed in roads nearby parking lot and information regarding car like colour and license plate recognition is captured and saved in the database [6]. S. V. Srikanth et.al, (2009) proposed a parking system which eliminates problems regarding finding vacant slot for parking. Author uses wireless technology to enhance parking efficiency [7]. Gongjun Yan et.al, (2008) proposed NOTICE based parking system. In this parking system, drivers can check and reserve the slot for parking. For security purposes encryption/decryption techniques are used. Simulation results are highly efficient [8]. Insop Song et.al, (2006) proposed FPGA based parking system using fuzzy logic controller (FLC). Reduction in computation time is its advantage. In this research work a robot car is made and tested in real environment using VHDL code. Design is simulated and tested on FPGA [10].

## 3. PROPOSED PARKING SYSTEM

The basic operation of the parking system is explained as: When a vehicle enters in the parking lot, LCD displays, if the space is available in parking lot or not. If the space is available then stepper motor rotates and door opens for vehicle entrance. RF module is used to transmit and receive slot availability information. According to RF Module's output, LED's glow. According to information driver can park the vehicle. Block diagram for the parking system is shown in figure 2. Host computer acts as control unit. Once host computer program the FPGA,

Identification and slot checking modules activate.

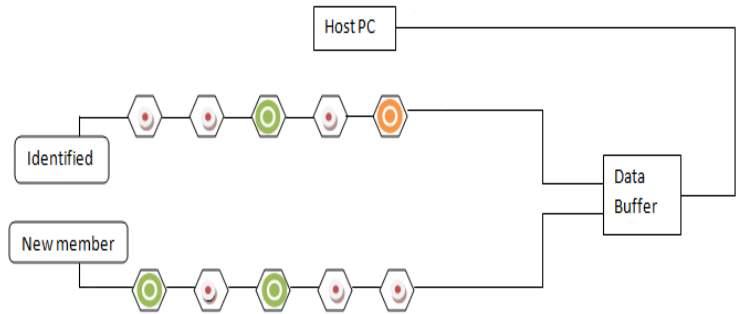

Figure 2: Block diagram of the parking system

## 3.1 Hardware implementation

LCD is of 16 pin configuration. $V_{ss}$ is ground pin. $V_{cc}$ is power supply pin. $V_{ee}$ is used to control the contrast. RS is register select pin. Command and data is select according to RS pin status. E is enable signal. DB0-DB7 are data lines. 15 and 16 pins are used to adjust backlight. Control and data lines of LCD assigned to FPGA are shown in figure 3.

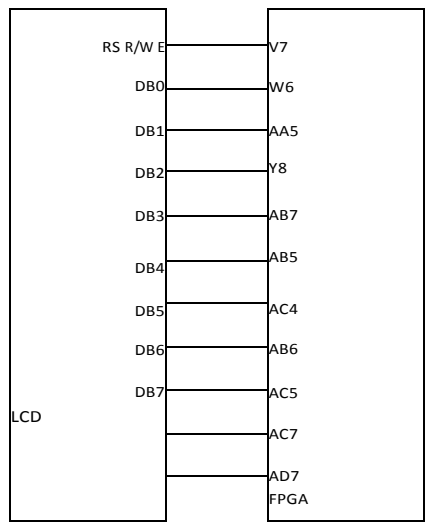

Figure 3(a): LCD interfacing with FPGA (Pin assignment)

Stepper motor interfaced with FPGA by using ULN 2003. Any I/O port of Development Board can be used for interfacing. Here, D I/O port of Development Board is used for interfacing with stepper motor. Pin assignment of FPGA, ULN 2003, and stepper motor is shown in figure 3(b).

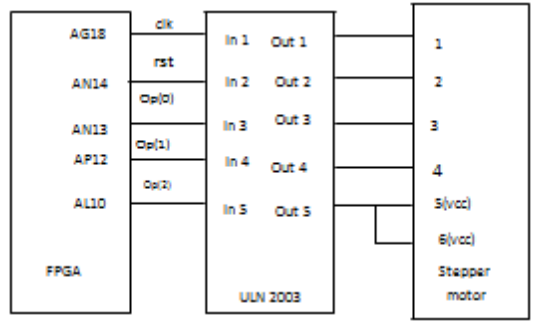

Figure 3(b): stepper motor interfacing with FPGA (Pin assignment)

RF Module: In this module IR trans-receiver pair is used to detect vehicle presence. IR sensors transmit slot's status to HT12E encoder. Encoder consists of 18 pin configuration. Parallel data convert into serial by encoder. Data is collected at data out pin of encoder. From data out pin data serially transmit towards RF transmitter. Data serially received at RF receiver. HT12 D decoder receiver's data from RF receiver, then converter back serial to parallel form. HT12D decoder's data pins are interfaced with Virtex 5 C I/O port signal pins.

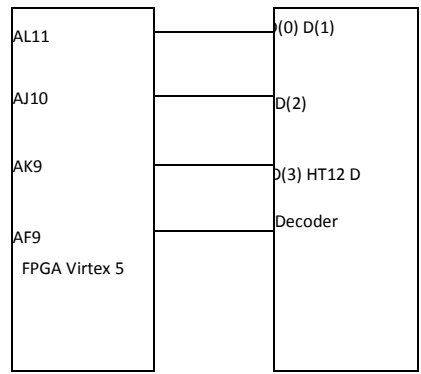

Figure 3(c): HT12D Decoder interfacing with FPGA (Pin assignment)

Platform: consisting of IR sensors, HT12E encoder, RF module, HT12D decoder and LEDs of FPGA are used to display results.
Software Module: Xilinx version 12.4 is used for VHDL coding.
Interfacing: In proposed system LCD, Stepper motor and RF models interfacing is done with FPGA.

## 3.2 Finite state machine for the operation of the system

In order to implement parking system a state diagram is constructed. As we know that the proposed system performs space availability, motor rotation for door opening, identification and slot availability information. The system includes:

Identification
Slot Selection

## 4. FPGA IMPLEMENTATION

### 4.1 Designing of parking system with State Machine Diagram

ASMD chart is Algorithm State Machine Diagram. This shows the working of parking system in the form of a flow chart. For the proposed model, oval shaped boxes are used to describe the output that depends upon past state as well as present input. The ASMD chart shown in figure 4 gives working of the car parking system.  At the entrance of parking area, LCD displays the status of parking system.  If space is available then LCD displays space available else LCD displays no space exit. According to space status motor rotates in clockwise direction. After that identification unit identifies the person. For  new member temporary card is allotted. After identification, slot status is checked. Status can be filled, empty or reserved. RF sensors are used in this process.

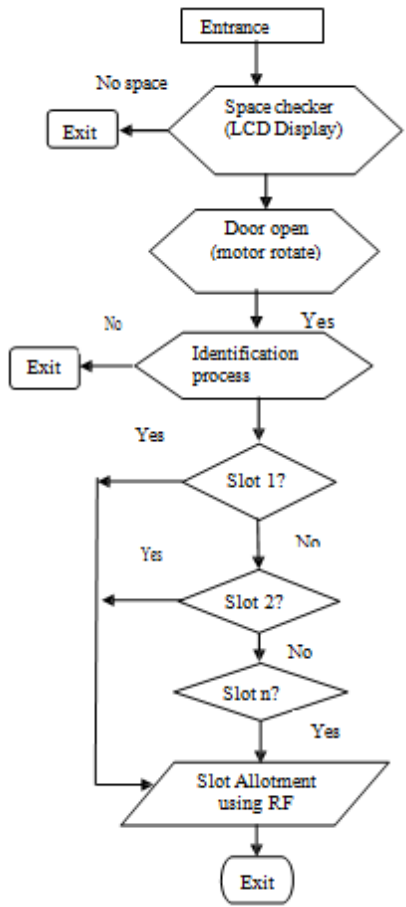

Figure 4: ASMD chart for parking system

## 4.2 Results

After space checking door will open with the help of stepper motor. Here in simulation clk pulse and reset is applied as an input. Cnt and clkd are signals. When reset goes high-to-low, stepper motor rotates. Simulation wave forms of stepper motor are shown in figure 4(a) clk is system clock signal. rst is control signal. D(7:0) are data lines. Z(3:0) is output signal of stepper motor. E is enable signal of LCD. rs is register select signal. rw is read/write control signal.

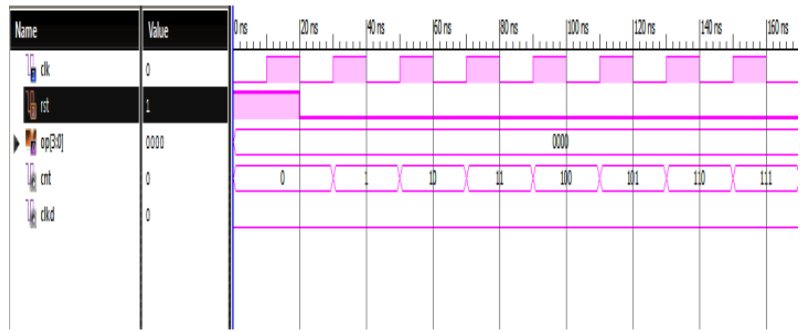

Figure 4(a): Simulation of stepper motor rotation

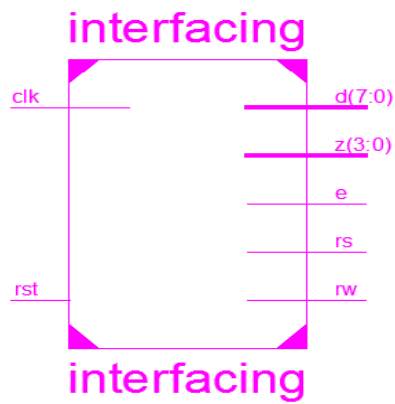

Figure 4(b) :RTL view of stepper motor and LCD interfacing

Figure above shows the RTL view of stepper motor and LCD interfacing.

When door opened, identification process starts. w, w1, w2, z, clk and reset are inputs. Out_1 is output. Current_state and next_state describes visitor is identified or a new member has come. Pr_st and nx_st shows person which is identified. Following simulation shows identification process:

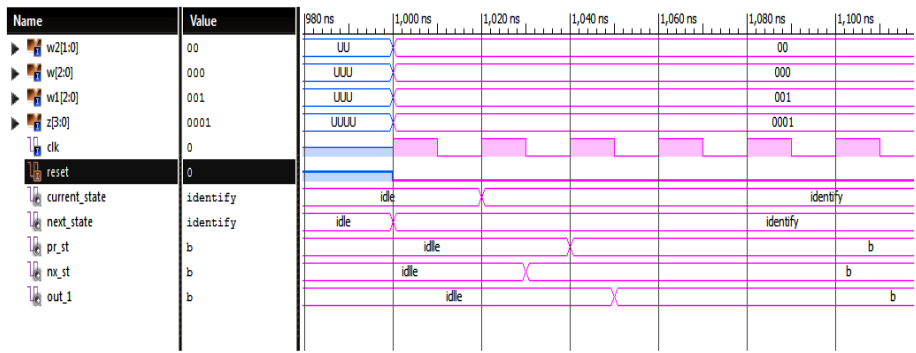

Figure 4(c): Simulation of identification module

After that slot checking procedure starts. Here w1, w2, w3, w, clk, reset are input signals. Led_slotallot and slotallot are output signals. When reset signal goes high-to-low, system comes out from idle state. According to input signals in following simulation slot 15 is available. Following simulation shows slot allotment feature.

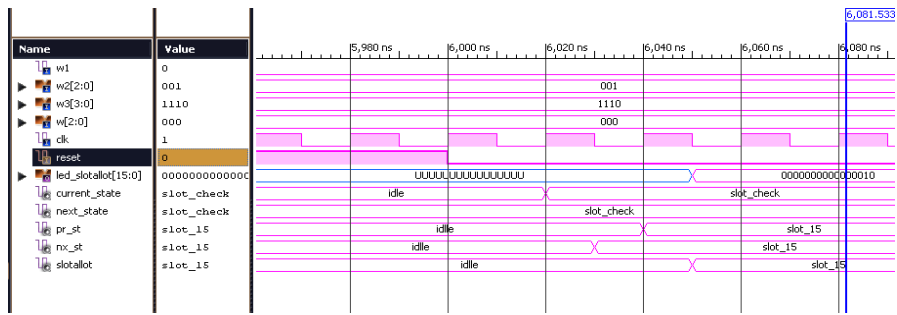

Figure 4(d): Simulation of slot allotment feature

Now identification and slot allotment modules are integrated. clk, w3, car_enter, reset, w4, fnd , a, w2 are input signals. Identified,new_member,fnd1,z, led,led_filled,led_reserv,cout are output signals. According to input signal, slot status is checked.

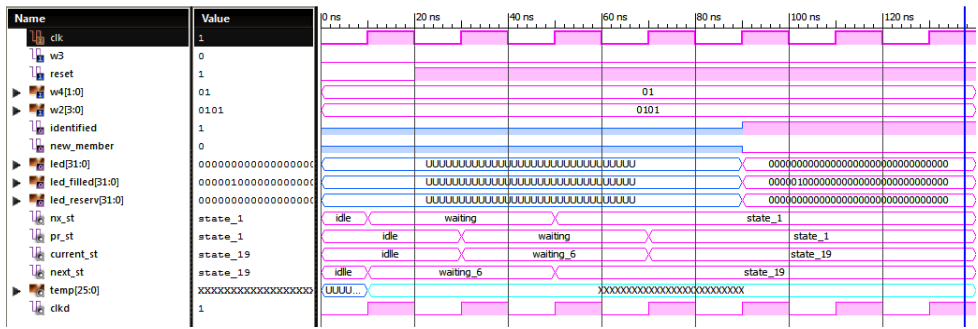

Figure 4(e): Simulation results of complete parking system

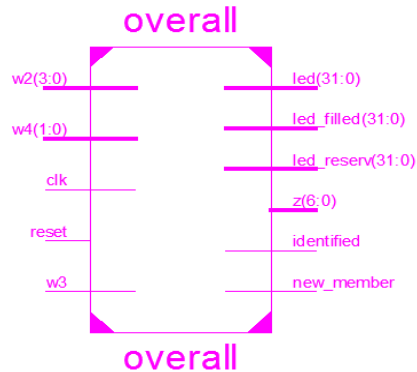

Figure 5: RTL view of parking system

Figure above shows the 32 slot involving RTL view parking system. W2, W3, W4 are input signals. Reset is control signal. Clk is system clock signal. Led, led_filled, led_reserv are output signals, which shows slot status. Identified and new_member are also output signals, which shows result of identification module.

## 5. CONCLUSION

The present FPGA based parking system is implemented using FSMs with the help of Xilinx ISE Design Suite 12.4. The design is verified on Virtex 5 FPGA kit. State machines increase productivity, reduces cost, and accelerates time to market. FPGA based parking system, gives fast response. The designed system can be used for many applications and can easily enhance the number of slot selections. Parking becomes easy by the use of Designed system.

## AUTHORS BIOGRAPHY

Balwinder Singh has obtained his Bachelor of Technology degree from National Institute of Technology, Jalandhar and Master of Technology degree from University Centre for Inst. & Microelectronics (UCIM), Punjab University, and Chandigarh in 2002 and 2004 respectively. He is currently serving as Senior Engineer in Centre for Development of Advanced Computing (CDAC), Mohali and is a part of the teaching faculty and also pursuing Phd from GNDU Amritsar.
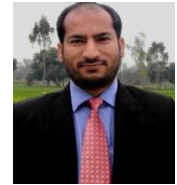
He has 8+ years of teaching experience to both undergraduate and postgraduate students. Singh has published three books and many papers in the International & National Journal and Conferences. His current interest includes Genetic algorithms, Low Power techniques, VLSI Design & Testing, and System on Chip.

Ramneet kaur has received the B.Tech. (Electronics and Communication Engineering) degree from the CTIEMT, Jalandhar affiliated to Punjab Technical University, Jalandhar in 2011, and presently she is doing M.Tech (VLSI design) degree from Centre for of Advanced Computing (CDAC), Mohali and working on her thesis work. Her area of interest is FPGA Implementation and VLSI Design.
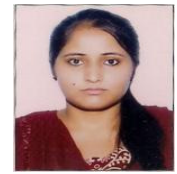